\documentclass[conference]{IEEEtran}
\IEEEoverridecommandlockouts
\usepackage{cite}
\usepackage{amsmath,amssymb,amsfonts}
\usepackage{algorithm,algorithmic}
\usepackage{array}
\usepackage{eqparbox}

\usepackage{graphicx}
\usepackage{textcomp}
\usepackage[table,xcdraw]{xcolor}
\usepackage{xcolor}
\usepackage{booktabs}
\def\BibTeX{{\rm B\kern-.05em{\sc i\kern-.025em b}\kern-.08em
    T\kern-.1667em\lower.7ex\hbox{E}\kern-.125emX}}

\newcommand{\fakeparagraph}[1]{\smallskip\noindent\textbf{#1.}}

\begin{document}

\title{A Blockchain-based Approach for Assessing Compliance with SLA-guaranteed IoT Services   \\
}

\author{\IEEEauthorblockN{Ali Alzubaidi\IEEEauthorrefmark{1}\IEEEauthorrefmark{2}, Karan Mitra\IEEEauthorrefmark{3}, Pankesh Patel\IEEEauthorrefmark{4} and Ellis Solaiman\IEEEauthorrefmark{5}} \IEEEauthorblockA{\IEEEauthorrefmark{1}Newcastle University, School of Computing, UK, Email: A.A.K.Alzubaidi2@newcastle.ac.uk}

\IEEEauthorblockA{\IEEEauthorrefmark{2}Umm Al-Qura University, Saudi Arabia, 
Email: aakzubaidi@IEEE.org}

\IEEEauthorblockA{\IEEEauthorrefmark{3}Luleå University of Technology, Sweden, Email: karan.mitra@ltu.se}

\IEEEauthorblockA{\IEEEauthorrefmark{4}Pandit Deendayal Petroleum University, India, Email: pankesh.patel@sot.pdpu.ac.in}

\IEEEauthorblockA{\IEEEauthorrefmark{5}Newcastle University, School of Computing, UK, ellis.solaiman@newcastle.ac.uk}}

\maketitle
\begin{abstract}
Within cloud-based internet of things (IoT) applications, typically cloud providers employ Service Level Agreements (SLAs) to ensure the quality of their provisioned services. Similar to any other contractual method, an SLA is not immune to breaches. Ideally, an SLA stipulates consequences (e.g. penalties) imposed on cloud providers when they fail to conform to SLA terms. The current practice assumes trust in service providers to acknowledge SLA breach incidents and executing associated consequences. Recently, the Blockchain paradigm has introduced compelling capabilities that may enable us to address SLA enforcement more elegantly. This paper proposes and implements a blockchain-based approach for assessing SLA compliance and enforcing consequences. It employs a diagnostic accuracy method for validating the dependability of the proposed solution. The paper also benchmarks Hyperledger Fabric to investigate its feasibility as an underlying blockchain infrastructure concerning latency and transaction success/fail rates.
\end{abstract}


\section{Introduction}
A Service Level Agreement (SLA) is a contracting method, commonly employed by cloud providers to ensure the quality of an X-as-a-Service (i.e. IaaS, PaaS, SaaS, etc.)~\cite{Faniyi2015}. It specifies rights and obligations of involved participants with relation to agreed Service Level Objectives (SLOs). An SLA declares to which level a Quality of Service (QoS) must be maintained, and what associated consequences (e.g. penalty, remedy, etc.) applied in case of SLA violation~\cite{Alqahtani2019}\cite{Alqahtani2018a}. 

In current practice of incident management, cloud providers promise to process incidents in good faith, assuring their consumers to impose SLA consequences on themselves. While intriguing, it is typically the consumer’s responsibility to report a service level degradation, supported by evidence deemed irrefutable by the service provider. This is usually a tedious process and manually handled \cite{Alzubaidi2019}. Most related enforcement studies, such as in \cite{Faniyi2015} and\cite{Mubeen2018}, assume trust in either service providers or trusted third parties. However, it can be inviting for some consumers to manipulate evidence to support violation incidents. On the other hand, providers may not react well to poorly formed claims, regardless of their validity \cite{Scheid2019}. In some scenarios, unresolved disputes have to be escalated to mediators or other jurisdiction means \cite{Rana2008}.

Recently, the rise of Blockchain technology invites revisiting existing solutions, where trust is taken for granted; Incident management is of no exception. The concept of smart contracts has enabled blockchain-based decentralised applications to serve various problem domains. Thus, we see an opportunity in exploiting Blockchain features to enable non-repudiable enforcement of SLA consequences.

We can list a set of advantages of shifting SLA enforcement to blockchain environment as follows: 

\begin{itemize}
    \item Improving incident management's procedures. Reported incidents can be automatically raised and instantly handled in a non-repudiable manner. 
     \item SLA terms are expressed in a logical format in the form of a smart contract.
    \item Records' immutability can help minimise dispute cases as well as the need for escalation.  
    \item Service providers can reduce the workforce size allocated for handling such tedious tasks. 

\end{itemize}

\subsection{Contribution}
The main contributions of this paper are: 

\begin{enumerate}[\IEEEsetlabelwidth{9)}]
    \item The paper proposes a decentralised approach for enforcing consequences of SLA violations. It models and implements a smart contract logic that addresses incidents processing, and automates decision-making on the compliance level of obligated providers with their offered SLA.
    \item It employs a validation method, known as Accuracy Diagnostic \cite{Simundic2009}, for examining the dependability of the modelled smart contract in every development iteration.
    \item The paper evaluates the feasibility of Hyperledger Fabric in terms of both the latency and transactions Success/fail rate, and provides a set of observations.
\end{enumerate}
The source code of our implementation is publicly available under GNU GPL V3.0 License on GitHub\footnote{https://github.com/aakzubaidi/MQTT-SLA-Blockchain-QoS-Enforcement}. We hope it forms a base that can help other interested researchers and industry alike to advance blockchain-based SLA management. \\

\fakeparagraph{Outline} The rest of the paper is organised as follows: First, section~\ref{sec:researchContext} elaborates on the research context, by setting an IoT scenario as a motivating example, and making use of a real SLA. It also briefly overviews Hyperledger Fabric, that we adopt as underlying blockchain infrastructure. In section~\ref{sec:Proosal}, we overview the proposed approach, and delve into modelling the proposed logic of enforcing SLA consequences. Section~\ref{sec:SimandVald} presents a simulation experiment on the feasibility of the approach. It also presents how we validate the smart contract for every development iteration. Finally, section~\ref{sec:PerformanceandResults} evaluates the performance of Hyperledger Fabric, aiming to assess its viability as an underlying blockchain infrastructure.

\section{Research context}
\label{sec:researchContext}
The IoT World Forum sets an Internet of Things (IoT) reference model, where components are organised in a multi-layered architecture \cite{Stallings2015}. It suggests a point where communication, processing, and storage are being centrally managed. Cloud services have been widely employed to serve such centric tasks in most IoT conceptualised architectures \cite{White2017}. In practice, cloud providers, such as Google IoT Platform\footnote{https://cloud.google.com/solutions/iot}, provide a set of pay-as-you-go services for registering devices, handling exchanged data through HTTP servers or MQTT brokers, processing data, providing analytical tools, and big-data storage means. Consumers (i.e. healthcare providers) may find it appealing to outsource such tasks to cloud providers since it alleviates several IT burdens, including scalability and maintenance. In return, cloud providers ensure a satisfactory service delivery through the concept of SLA.

 \begin{figure}[htbp]
 \centering
     {\includegraphics [width=0.3\textwidth]{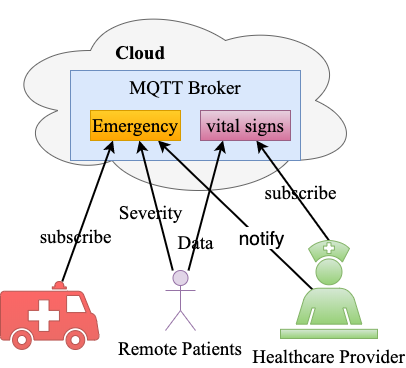}}
     \caption{Example IoT healthcare scenario employing MQTT for data exchange}
     \label{MQTTScenario}
 \end{figure}

 \subsection{Motivating Scenario}
 \label{sec:scenario}
Message Queuing Telemetry Transport (MQTT)\footnote{http://mqtt.org/} is an asynchronous data exchange protocol based on publish/subscribe model. Since HTTP is document-oriented, it is less favourable when it comes to constraint resources. Therefore, MQTT, being data-oriented, is a popular data exchange protocol for IoT applications purposes. The mechanism of MQTT relies on the concept of topics, to which authorised entities can publish or subscribe. MQTT brokers orchestrate the data exchanged between publishers and subscribers. Because of the popularity of MQTT, several cloud providers offer MQTT brokers as a service. 

To elaborate, consider a Telemedicine example where a healthcare provider hires cloud services. These services depend on a MQTT broker, which is used for communication and data exchange. For example, Figure~\ref{MQTTScenario} depicts an IoT healthcare application that enables monitoring patients remotely and reacting in case of emergency. In this simplified scenario, three major entities communicate over the MQTT protocol, which are patients, the healthcare provider, and an ambulance department. These entities publish/subscribe to topics of their interest. For instance, health data can be collected by sensors/devices, and then published to a topic called \textit{vital signs}. If a patient's condition becomes severe, both the healthcare provider and the ambulance department will be notified through the Emergency topic.

 \subsection{SLA Example}
    \label{sec:slacase}
Cloud providers cope with IoT requirements by offering enabling services such as MQTT servers. For IoT scenarios such as the one presented in section~\ref{sec:scenario}, cloud providers guarantee the quality of MQTT servers using the concept of SLA. For this paper, we use the GCP (Google Cloud Platform) SLA \footnote{https://cloud.google.com/iot/sla} for modelling our approach. The SLA covers a component for bridging between cloud services and IoT devices using MQTT. The GCP SLA ensures compensating consumers when a set of valid MQTT requests results in accidental device connections. The SLA holds the cloud provider accountable for consequences when the error rate exceeds 10\%. For simplicity, we consider the error rate as the total accidental device disconnections divided by the total number of valid MQTT requests. For clarity, we formally define the error rate as in Equation~\ref{error-rate}, where $F$ stands for MQTT Fail Requests and $V$ stands for MQTT Valid Requests.
 \begin{equation} \label{error-rate}
 Error Rate = (\dfrac{\sum\limits_{i=1}^{n}F}{\sum\limits_{i=1}^{n}V}) \times 100
 \end{equation}

Within this paper we consider the GCP SLA as a representative example, and base our approach on it for assessing provider compliance, and for enforcing the violation consequences (when the error rate exceeds the 10\% threshold).

\subsection{Service Monitoring}
Service monitoring is vital for enforcing satisfactory SLA compliance \cite{Rana2008}. Examples of monitoring/reporting tools are those surveyed in \cite{Mubeen2018}\cite{Hussain2014}, which can be classified into proactive or reactive categories. The former predicts potential failures while the latter identifies incidents that are already in place. Ideally, cloud providers would apply proactive measures to prevent violation in the first place. While proactive violation methods present an exciting class of problems, we still need to discuss the aftermath of SLA violations \cite{Hussain2014}. This paper is limited to reactive monitoring methods, which we employ for incidents identification.

\subsection{Hyperledger Fabric as Enabling Technology}
SLA as any other contracting method, is prone to trust and enforcement issues. Whenever trust is a burden, Blockchain present itself as a potential underlying infrastructure. Blockchain holds appealing principles including, but not limited to, immutability, decentralised computation, and a distributed shared ledger controlled by a consensus mechanism. The concept of a smart contract allows applications to benefit from these features, forming what is commonly referred to as Decentralised Applications (DAPPs) \cite{Cai2018}. We see an opportunity in taking advantage of Blockchain to serve SLA monitoring and enforcement purposes. This does not only address trust issues \cite{Scheid2019} but also expedites and automates associated processes such as handling violations and decision making. It also helps reduce the need for manual processing and human intervention.

We select Hyperledger Fabric (hereafter HLF) \cite{Androulaki2018} for realising the proposed blockchain-based SLA enforcement. HLF is classified as a consortium blockchain, composed of per-identified validating nodes. The permissioned nature of HLF enables adopting plugable consensus protocols lighter than Proof-of-work (PoW). The latest official version of HLF (v1.4.6 as of today) officially supports various Crash-Tolerant protocols as consensus mechanisms such as Raft \cite{Ongaro2014} and Kafka/ZooKeeper. HLF supports general-purpose programming languages for representing the business logic as a smart contract. Supported languages include Golang, Javascript, and Java. Since all nodes are authenticated and committed for execution and validation tasks, there is no need for any execution fee, as it would be the case with public blockchain technologies. All transactions, whether successful or not, are immutably stored on the ledger for auditing purposes. Further details on the transaction flow can be found in \cite{Androulaki2018}.


\section {Blockchain-based Consequences Enforcement}
\label{sec:Proosal}
This section provides an overview on our approach, and then models a smart contract logic that enforces SLA consequences as per the GCP SLA example presented in section~\ref{sec:slacase}. 

 \begin{figure}[htbp]
 \centering
    {\includegraphics [width=0.45\textwidth]{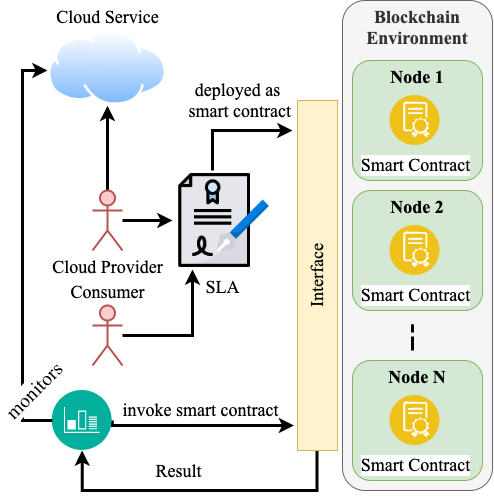}}
    \caption{Overview on the proposed Blockchain-based SLA enforcement}
    \label{GeneralArchetecture}
\end{figure}

\subsection{Overview}
As any contractual method, SLA is susceptible to breaches. In the current practice, SLA violations must be acknowledged by obligated providers in order to execute SLA consequences. Figure~\ref{GeneralArchetecture}, suggests revisiting the current trust model, by shifting some such critical tasks from obligated providers to executable contracts operating in a non-repudiable fashion. In particular, We shift three main incident management tasks from obligated providers, which are Incidents processing, SLA compliance assessment, and SLA consequences enforcement. Blockchain-based Smart contracts can realise this aim beyond the influence of any single entity. 

We highlight essential considerations for realising our approach as follows:
\begin{itemize}
    \item The SLA document in place: We study the SLA content, and extract key elements which include but not limited to:
    \begin{itemize}
        \item Participants: their rights and obligations.
        \item QoS metrics: their definition, measurement, and violation consequences.
    \end{itemize}
    \item Executable actions in the SLA must be represented in the smart contract, for example compliance assessment and consequences enforcement. The smart contract is then deployed to the blockchain network.
    \item SLA awareness within Blockchain: smart contracts need to have the necessary SLA awareness in order to execute relevant actions.
    \item Monitoring agent awareness: we have to consider the fact that smart contracts are supposed to be terminable and deterministic \cite{Wang2019}. Smart contracts are not optimal for conducting endless activities such as monitoring. Thus, external monitoring/reporting means have to be in place to help smart contracts in forming a decision on the compliance level of obligated providers. For illustration, consider the GCP SLA definition of MQTT error rate. Monitoring agents should collect metrics related to the performance of the MQTT server such as up-time, the number of received/sent messages. These metrics help the smart contract to form a decision on the compliance status of the obligated provider. That is, monitoring agents must submit any identified incidents to the smart contract.
\end{itemize}

The rest of the paper delves into exploiting the smart contract for processing monitoring logs, and assessing the conformance of cloud providers to their offered SLAs.

\subsection{Smart Contract Logic Modelling}
\label{sec:enforcement}

We use the concept of smart contracts to automate decision-making on the compliance level of obligated providers (Cloud provider), beyond the control of centralised authorities. The decision making is formed with the aid of monitoring tools. 

\begin{figure}[htbp]
 \centering
    {\includegraphics [width=0.45\textwidth]{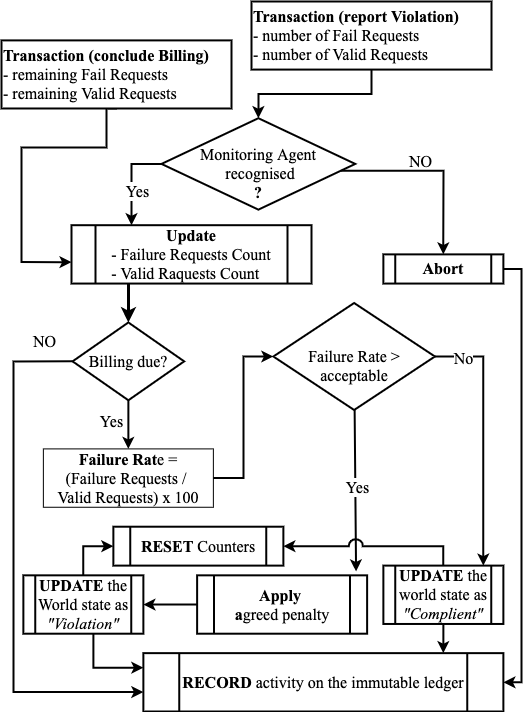}}
    \caption{Modelling the enforcement logic as a smart contract}
    \label{Mqtt-smartcontact}
\end{figure}

Figure~\ref{Mqtt-smartcontact} illustrates the proposed enforcement logic modelled as a smart contract. Two stages take place in every billing cycle (every month, for example). In the first stage, the smart contract waits for failure events, as per defined in the SLA. For that, there is an interface for authorised monitoring agents to invoke whenever there is an incident, as shown in Figure~\ref{GeneralArchetecture}. For every transaction, the smart contract accommodates incidents by updating the total number of two elements, which are \textit{Valid} and \textit{Fail} MQTT requests. These two elements are essential for calculating the error rate. It is worth mentioning that update operations are executed at the state database provided by HLF, and not in the the immutable ledger. However, every update operation is backed-up with an immutably stored transaction for auditing purposes.

The second stage takes place at the end of the billing cycle. At this stage, the smart contract decides on the compliance status of the obligated provider. The decision is mainly based on \textit{Error Rate}, calculated as per Equation~\ref{error-rate}. Then, it examines the calculated error rate against the threshold stipulated in the SLA in place; say 10\% for example. If the smart contract concludes that the obligated provider has fulfilled its promise by not reaching the stipulated threshold, then the performance will be marked as \textit{compliant}. Otherwise, it marks the performance as \textit{Violation} and applies the agreed penalty. Either way, the result will be immutably recorded on the ledger for reference purposes. The count of both \textit{Valid Requests} and \textit{Failure Requests} are then reset for the next billing cycle.

\section{Simulation and Validation}
\label{sec:SimandVald}
This section attests the feasibility of implementing our approach as a smart contract. It also describes how we validate the dependability of the proposed approach. The smart contract, simulation, and the validation process are implemented in Java programming language and available as open source in GitHub\footnote{https://github.com/aakzubaidi/MQTT-SLA-Blockchain-QoS-Enforcement}.

\subsection{Simulation and Implementation}
\label{sec:simulation}

\begin{figure}[htbp]
 	\centering
	{\includegraphics [width=0.35\textwidth]{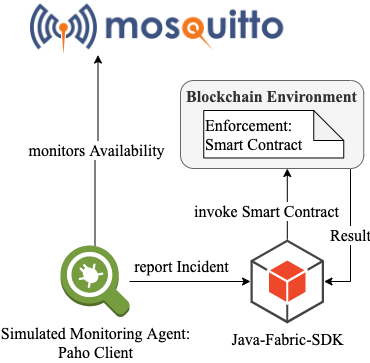}}
	\caption{Simulation of MQTT Broker and Monitoring agent}
	\label{Mqtt-experiment}
\end{figure}

The purpose of the simulation is to experiment with the feasibility of the proposed approach. Figure~\ref{Mqtt-experiment} illustrates the outlook of the implemented components. First, we employ an MQTT broker using \textit{Eclipse Mosquitto}\footnote{https://mosquitto.org/}. Second, we implement a simple behaviour of a monitoring agent using \textit{Eclipse Paho}\footnote{https://www.eclipse.org/paho/}. Third, the implemented smart contract, discussed in section~\ref{sec:enforcement}, is then deployed and instantiated using IBM Blockchain Platform~\footnote{https://marketplace.visualstudio.com/items?itemName=IBMBlockchain.ibm-blockchain-platform}. We also developed a bridge using Fabric Java SDK\footnote{https://github.com/hyperledger/fabric-sdk-java}, which represents the interface between the blockchain network and external entities (e.g. monitoring agents) to enable configuring all necessary elements (such as identity, access level, crypto materials, etc.).

The logic of our simulation and testing is as follows. The monitoring agent continuously publishes testing samples to the MQTT Broker (Mosquitto). The goal of the monitoring agent is to observe Valid and Fail MQTT requests. The monitoring agent uses exception handlers and error codes provided by~\textit{Eclipse Paho} to identify errors and reason about them. 
We deliberately cause the MQTT broker downtime to simulate cloud failure by simply shutting down the MQTT broker (Mosquitto). Once the monitoring agent detects an availability issue (downtime), it submits a transaction to the smart contract, reporting latest \textit{Valid} and \textit{Fail} MQTT request. The smart contract records incidents as well as valid MQTT requests, and will keep accumulating incidents until the error rate reaches 10\%. Then we test the the smart contract's ability to mark the cloud provider's performance as \textit{violation}. Overall, by conducting the above, our proposed approach proves to work as expected. Following, we describe our validation approach.


\subsection{Decision Accuracy Validation}

To ensure the reliability of our approach we consider two facts. First, the enforcement logic is an assessment tool in nature, which decides on the compliance level of obligated providers. Therefore, the dependability of the decision logic has to be validated \cite{Magazzeni2017}. Second, smart contracts operate beyond the control of a single authority. Therefore, rectifying a logical error in decentralised applications is not as straightforward as it would be in traditional applications \cite{Christidis2016}. Thus, this section presents how we validate the dependability of the enforcement logic for every development iteration on the smart contract. It examines the two stages of the enforcement life-cycle, which are the incident reporting as in Algorithm~\ref{Algo:CorrectValues} and compliance assessment as in Algorithm~\ref{Algo:DiagnosticAccuracy}. For every development iteration on the smart contract, we execute these validation algorithms using Junit testing framework.

\subsubsection{\textbf{Stage 1: Incidents Reporting}}
\label{sec:Incidents-Reporting}
The decision logic is based on Equation~\ref{error-rate}, which is composed of two elements: \textit{MQTT Fail Requests} and \textit{MQTT Valid Requests}. Algorithm~\ref{Algo:CorrectValues} investigates the smart contract ability to update the values of these elements correctly. It simulates a monitoring agent that repeatedly reports incidents to the smart contract. For each incident, the simulated monitoring agent submits MQTT Fail Requests $F$, and MQTT Valid Requests $V$. The smart contract will record this incident and update values. The algorithm queries the current count of both \textit{Valid} and \textit{Fail MQTT requests}, $V\_count$ and $F\_count$, and assert that they are appropriately updated.
\begin{algorithm}
	\caption{Validating correct values update}
	\begin{algorithmic}[1]
		\label{Algo:CorrectValues}
		\renewcommand{\algorithmicrequire}{\textbf{Input: }}
		\renewcommand{\algorithmicensure}{\textbf{Output:}}
		\REQUIRE  $F$, $V$
		\ENSURE  $F\_count$ and $V\_count$ are correctly updated 
		\FOR {$i \leftarrow 1$ to $Transactions\_count$}
		\STATE \textbf{Invoke} Report\_Violation ($F$, $V$)
		\STATE \textbf{Wait} for transaction resolution
		\STATE \textbf{Query} ($F\_count$ , $V\_count$)
		\IF{$F\_count$ OR $V\_count$ NOT correctly updated}
		\STATE \textbf{Terminate} with error
		\ENDIF
		\ENDFOR
		\RETURN $V\_count$ , $F\_count$
	\end{algorithmic} 
\end{algorithm}

We always assume the worst-case scenario, where the MQTT broker exhibits malfunction behaviour lasting for the entire 30 days. We adopt the default reporting interval employed by Google Stackdriver monitoring tool \footnote{https://cloud.google.com/monitoring/api/metrics\_gcp}, which is every 60 seconds. Thus, the monitoring agent will keep triggering the smart contract every minute. This means the smart contract should receive 43200 transactions by the end of the month. Therefore, we iterate for 43200 times, such that there are new Valid $V$ and Fail $F$ MQTT requests for every iteration. Thus, by the end of algorithm execution, $V\_count$ should equals $V\times43200$ and $F\_count$ should equals $F\times43200$.

\subsubsection{\textbf{Stage 2: Compliance Assessment}}
\label{sec:DiagnosticAccurDec}
At the end of every billing cycle (assume 30 days), the smart contract assesses the compliance level of obligated providers based on the agreed \textit{Error Rate Threshold}, For example 10\%. It calculates the Error rate as per Equation~\ref{error-rate}, and compares it with the agreed \textit{Error Rate Threshold} in the SLA.
To examine decision accuracy we prepared a set of cases already known to us to be either \textit{Violation} or \textit{Compliant}. The Violation group has 50\% of these cases, which exceed the 10\% Error Rate Threshold. The Compliant group has the other half of the cases, which does not exceed the 10\% Error Rate Threshold.
Given the importance of this assessment task we employ rigorous testing that validates the decision made by the smart contact. We rely on a method called \textit{Diagnostic Accuracy}  \cite{Simundic2009}, which employs a set of measures and calculations based on 2X2 table, as shown in Table~\ref{table:DAtable}. The table is composed of 4 elements which are True Positive (TP), True Negative (TN) False Positive (FP), False Negative (FN). Let us assume an SLA that stipulates a service delivery with an Error Rate that does not exceeds 10\%; Otherwise, it is considered a violation case. Then, we consider \textit{TP} is the count of the cases that is correctly classified to be breaching this terms (the 10\% threshold). Therefore, \textit{TN} is the count of the cases that are correctly classified to be compliant and does not exceed the 10\% threshold. \textit{FP} and \textit{FN} are the cases where the smart contract incorrectly identify cases as Violation or Compliant; respectively.

\begin{table}[ht] \renewcommand{\arraystretch}{1.3} \caption{Diagnostic Accuracy 2x2 Table} \label{table:DAtable}
	\centering
	\begin{tabular}{c|c|c|}
		\cline{2-3}
		\multicolumn{1}{l|}{}          & Violation                                                     & Complaint                                                      \\ \hline
		\multicolumn{1}{|c|}{Positive} & \begin{tabular}[c]{@{}c@{}}True Positive \\ (TP)\end{tabular} & \begin{tabular}[c]{@{}c@{}}False Positive \\ (FP)\end{tabular} \\ \hline
		\multicolumn{1}{|c|}{Negative} & \begin{tabular}[c]{@{}c@{}}False Negative\\ (FN)\end{tabular} & \begin{tabular}[c]{@{}c@{}}True Negative \\ (TN)\end{tabular}  \\ \hline
	\end{tabular}
\end{table}

Based on the Diagnostic Accuracy table, we select and define the following measurements:
\begin{itemize}
	\item \textit{Sensitivity:} the proportion of violation cases correctly classified by the smart contract; calculated as follows: 
	
	\begin{equation} \label{EQU:senstivity}
	Sensitivity = \dfrac{TP}{TP+FN} 
	\end{equation}
	
	\item \textit{Specificity:} the proportion of compliant cases correctly classified by the smart contract; calculated as follows:

	\begin{equation} \label{EQU:Specificity}
	 Specificity= \dfrac{TN}{TN+FP}
	 \end{equation}
	 
	 \item \textit{Positive Predictive Value (PPV):} The probability of accurate decision on violation cases; calculated as follows: 
	 
	 	\begin{equation} \label{EQU:PPV}
	 PPV= (\dfrac{TP}{TP+FP}) \times 100
	 \end{equation}
	 
	 	 \item \textit{Negative Predictive Value (NPV):} The probability of accurate decision on compliant cases; calculated as follows: 
	 
	 \begin{equation} \label{EQU:NPV}
	 NPV= (\dfrac{TN}{TN+FN}) \times 100
	 \end{equation}

\end{itemize}

The goal is to examine the smart contract ability to correctly classify the prepared cases into their corresponding groups; either \textit{Violation} or \textit{Compliance}. Algorithm~\ref{Algo:DiagnosticAccuracy} illustrates the conduct of the Diagnostic Accuracy method. First, We feed all prepared sets of cases into the smart contract. We fill in table~\ref{table:DAtable} according to the outcome of the smart contract. That is, for each case we classify the result of the smart contract decision on every case to be one of the listed categories (TP, TN, FP, or FN). We conduct the calculation of \textit{Sensitivity}, \textit{Specificity}, \textit{PPV} and \textit{NPV} to find out about accuracy of the compliance assessment. We implemented this validation method (Diagnostic Accuracy) using Junit testing framework, and executed it for every development iteration on the smart contract. Results are shown in table~\ref{table:DAresult}.

\begin{table}[ht] \renewcommand{\arraystretch}{1.3} \caption{Optimum results of the Diagnostic Accuracy method} \label{table:DAresult}
	\centering
	\begin{tabular}{|c|c|c|c|}
		\hline
		Sensitivity & Specificity & PPV   & NPV   \\ \hline
		100\%       & 100\%       & 100\% & 100\% \\ \hline
	\end{tabular}
\end{table}

   \begin{algorithm}
   	
   	\caption{Conducting The Diagnostic Accuracy Method}
    \label{Algo:DiagnosticAccuracy}
   	\begin{algorithmic}[1]
   		\renewcommand{\algorithmicrequire}{\textbf{Input: }}
   		\renewcommand{\algorithmicensure}{\textbf{Output:}}
   		\REQUIRE  $TEST\_CASES$
   		\ENSURE  $Sensitivity$ , $Specificity$ , $PPV$ , $NPV$
   		\STATE $C = \{c: c \leq threshold\}$    \COMMENT{Compliant cases}
   		\STATE $V = \{v: v > threshold \}$   \COMMENT{Violation cases}
   		\STATE $TEST\_CASES = \{C \cup V\}$
   		\FORALL {$tc \in TEST\_CASES$}
   		\STATE \textbf{assessCompliance ($tc$)} \COMMENT{invoke the smart contract}
   		\STATE \textbf{determine} whether decision is $TP, TN, FP or FN$
   		\IF{TP}
   		\STATE $TP\_count++$
   		\ELSE
   		   	\IF {$TN$}
   			\STATE $TN\_count++$
   			\ENDIF
   		\ELSE
   			\IF {$FT$}
   			\STATE $FT\_count++$
   			\ENDIF
		\ELSE
			\IF {$FN$}
			\STATE $FN\_count++$
			\ENDIF
   		\ENDIF
   		\ENDFOR
   		\STATE \textbf{Calculate} $Sensitivity$ , $Specificity$ , $PPV$ , $NPV$

   	\end{algorithmic} 
   \end{algorithm}


\section{performance Evaluation and Results}
\label{sec:PerformanceandResults}
HLF has proven to perform well for several scenarios. There exists a set of studies on HLF performance such as those in \cite{Androulaki2018} \cite{Thakkar2018} \cite{Kuzlu2019} \cite{Gorenflo2019} which are based on Kafka/Zookeeper protocol and \cite{Dinh2017a} which is based on PBFT. To the best of our knowledge, there is no performance benchmarking coping with the latest development on HLF, where Raft protocol \cite{Ongaro2014} has been officially recommended as a consensus protocol. In this paper, we adopt Raft, and thus we focus on the latest official HLF version 1.4.6.

The previous section looked into the smart contract logic dependability. However, we still need to confirm the feasibility of HLF as an underlying Blockchain technology. Therefore, this section investigates how HLF network copes with consecutive incidents (assuming a very poorly performing cloud provider). In particular, we look into HLF's latency and transaction success/fail rate.

\subsection{Experimental Setup}
\label{sec:performance}
We focus on the most demanding functionality in our modelled smart contract, which is processing the received incidents, as described in Algorithm~\ref{Algo:CorrectValues}. At the infrastructure level, Table~\ref {table:HLFConfig} illustrates both, the testing environment and HLF configuration.

\begin{table}[ht] \renewcommand{\arraystretch}{1.3} \caption{Hyperledger Fabric (HLF) network deployment configuration} \label{table:HLFConfig}
	\centering
	\begin{tabular}{|l|l|}
		\hline
		\rowcolor[HTML]{EFEFEF} 
		\multicolumn{1}{|c|}{\cellcolor[HTML]{EFEFEF}\textbf{Factor}} & \multicolumn{1}{c|}{\cellcolor[HTML]{EFEFEF}\textbf{Settings}}                                                                                                                                                           \\ \hline
		Host Machine                                                  & \begin{tabular}[c]{@{}l@{}}Local running MacOS Catalina OS, \\ 2.9GHz Dual-core intel Core i5 CPU, \\ 2GB LPDDDR3 Memory Ram.\end{tabular}                                                                               \\ \hline
		Containerization                                              & \begin{tabular}[c]{@{}l@{}}Docker version 2.2, Engine version 19.3.5.\\ Allocated Resources: \\ CPUs: 3, Memory: 7GB, Swap: 3GB\end{tabular}                                                                             \\ \hline
		 Network Topology                                   & \begin{tabular}[c]{@{}l@{}}HLF Version 1.4.6, \\ 2 Organisations,\\ 4 committing peers, \\ 5 orderers, each one in a separate container. \\ One dedicated channel, \\ LevelDB is used as a ledger database.\end{tabular} \\ \hline
		Consensus Protocol                                            & Raft Protocol                                                                                                                                                                                                            \\ \hline
		Endorsement Policy                                            & \begin{tabular}[c]{@{}l@{}}One peer of each organisation\end{tabular}                                                                                                                     \\ \hline                                                                
		\rowcolor[HTML]{EFEFEF} 
		\multicolumn{2}{|c|}{\cellcolor[HTML]{EFEFEF}\textbf{Chaincode Settings}}                                                                                                                                                                                                                \\ \hline
		Execution Timeout                                             & 60 seconds                                                                                                                                                                                                               \\ \hline
		Programming Language                                          & Java                                                                                                                                                                                                                     \\ \hline
		Logging                                                       & Enabled                                                                                                                                                                                                                  \\ \hline
	\end{tabular}
\end{table}

We employ a modular blockchain benchmarking project under Hyperledger Umbrella, called \textit{Hyperledger Caliper} \footnote{https://hyperledger.github.io/caliper/v0.3/getting-started/} for deploying the smart contract, simulating the behaviour of monitoring agents, and benchmarking.

Table~\ref{table:caliperSet} summarises the experimental settings of 4 different test cases. They are all similar in terms of benchmarking settings, where we denote similarity as $\sim$ to indicate no change in the settings. However, these test cases are different in terms block batching configurations. For this experiment, we assume consecutive transactions. Therefore, we aim to figure out a suitable block batching configuration that avoids concurrency issues while maintaining a reasonable throughput and latency. We employed two factors for defining the readiness of a block for validation, which are: \textit{timeout}, and maximum allowed number of \textit{transactions per block}; whichever occurs first. These two factors vary in every testing, seeking the best performance possible.

\begin{table}[ht] \renewcommand{\arraystretch}{1.3} \caption{Experimental settings (Hyperledger Caliper)} \label{table:caliperSet}
	\centering
\begin{tabular}{|c|c|c|c|c|}
	\hline
	\rowcolor[HTML]{EFEFEF} 
	\textbf{Facet}                                                                     & \textbf{Test 1}                                                                                                                  & \textbf{Test 2} & \textbf{Test 3} & \textbf{Test 4} \\ \hline
	\begin{tabular}[c]{@{}c@{}}Function\\ under test\end{tabular}                      & \begin{tabular}[c]{@{}c@{}}Report Violation \\ (Asset update)\end{tabular}                                                       & $\sim$ & $\sim$ & $\sim$          \\ \hline
	\begin{tabular}[c]{@{}c@{}}total workers \\ (client thread)\end{tabular}           & 1 worker                                                                                                                         & $\sim$ & $\sim$ & $\sim$          \\ \hline
	\begin{tabular}[c]{@{}c@{}}control rate:\\ Fixed Rate:\end{tabular}                & \begin{tabular}[c]{@{}c@{}}300 \\ transactions\end{tabular}                                                                      & $\sim$ & $\sim$ & $\sim$          \\ \hline
	\begin{tabular}[c]{@{}c@{}}Sent Transactions \\ (per second)\end{tabular}          & 1 Tps                                                                                                                            & $\sim$ & $\sim$ & $\sim$          \\ \hline
	\begin{tabular}[c]{@{}c@{}}Execution \\ Timeout\end{tabular}                       & \multicolumn{1}{l|}{\begin{tabular}[c]{@{}l@{}}30 seconds for:\\ - Chaincode engine\\ - Worker\\ - Caliper runtime\end{tabular}} & $\sim$ & $\sim$ & $\sim$          \\ \hline
	\rowcolor[HTML]{C0C0C0} 
	\multicolumn{5}{|c|}{\cellcolor[HTML]{C0C0C0}\textbf{Fabric Block Batching Configuration}}                                                                                                                                                                         \\ \hline
	\begin{tabular}[c]{@{}c@{}}Block Batching \\ Timeout\\ (milliseconds)\end{tabular} & 1000                                                                                                                             & 500    & 1000   & 500             \\ \hline
	\begin{tabular}[c]{@{}c@{}}Transactions \\ per block\end{tabular}                  & 10                                                                                                                               & 10     & 1      & 1               \\ \hline
\end{tabular}
\end{table}

Our testing plan is that, we set all parameters in Table~\ref{table:caliperSet} to the minimal values possible, and then we scale gradually until we reach a bottleneck. For example, we started from 1 worker sending only one transaction. However, we encountered conflicting transactions for all different test cases as reported and discussed below. So we limited the number of workers to 1 for all four test cases and be more focused on Block batching configuration. We started with 1 second timeout and 10 transactions per seconds in test case 1. Then, we went to see how HLF would perform with lower values in the rest of test cases.

\subsection{Results and Observations}

\subsubsection{\textbf{Latency}}
All four test cases in Table~\ref{table:DAresult} reveal that HLF latency, measured in seconds, is generally acceptable given the limited resources of the host machine. We observe no major difference in average Latency, which exhibits a relatively similar behaviour. We can see from case 4 that, these is a clear fluctuation between minimum and maximum latency. We can attribute this to the high rate of commits on the ledger across the network \cite{Sukhwani2018}, resulted by generating a block every half second milliseconds. This leads many transactions to miss the chance of being included in the current block, and thus wait for the next one.

\begin{table}[ht] \renewcommand{\arraystretch}{1.3} \caption{Results of Latency and Transactions success rate} \label{table:results}
	\centering
\begin{tabular}{c|c|c|c|c|}
	& \cellcolor[HTML]{EFEFEF}\textbf{T1} & \cellcolor[HTML]{EFEFEF}\textbf{T2} & \cellcolor[HTML]{EFEFEF}\textbf{T3} & \cellcolor[HTML]{EFEFEF}\textbf{T4} \\ \hline
	\cellcolor[HTML]{EFEFEF}\textbf{Success}     & 298                                 & 297                                 & 299                                 & 295                                 \\ \hline
	\cellcolor[HTML]{EFEFEF}\textbf{Fail}        & 2                                   & 3                                   & 1                                   & 5                                   \\ \hline
	\cellcolor[HTML]{EFEFEF}\textbf{Max Latency (s)} & 1.22                                & 1.54                                & 1.43                                & 1.68                                \\ \hline
	\cellcolor[HTML]{EFEFEF}\textbf{Avg Latency (s)} & 0.77                                & 0.77                                & 0.69                                & 0.78                                \\ \hline
	\cellcolor[HTML]{EFEFEF}\textbf{Min Latency (s)} & 0.68                                & 0.68                                & 0.77                                & 0.34                                \\ \hline
\end{tabular}
\end{table}

\subsubsection{\textbf{Success/Fail Rates}}
\label{sec:failSucess}
We have to recognise the fact that HLF employs an optimistic locking mechanism, namely: Multi-Version Concurrency Control (MVCC) to prevent a known blockchain problem called double-spending problem \cite{Chung2019}. This mechanism caused all test cases to experience transactions failures due to conflict in MVCC Read-Write sets. The MVCC conflict is triggered because HLF experienced consecutive transaction reporting incidents to the smart contract. The evaluation results reveal that, test case 3, (1~Tps and 1s~timeout) experiences the least transactions failures. The worst performance is observed in test case 4, where block batching is set to the minimum possible (1~Tps and 500ms~timeout).

\subsection{Observation and Remarks}
While MVCC has proven to work well for some scenarios such as money transfer, it is not the case for demanding applications. For example, it is abnormal for a money transfer application to receive frequent updates on the same account within a few seconds. However, In our case, we ideally expect consecutive transactions from monitoring agents. Nevertheless, the lock mechanism causes some monitoring transactions to experience MVCC conflicting Read-Write sets, represented as (key, value version). When a transaction tries to update a key, it acquires the latest version of that key. Since our scenario assumes consecutive monitoring transactions, failure can accrue when transactions carry out update operations based on an obsolete key version. This situation can happen when an unsettled update transaction (not committed yet to the ledger) finally manages to be committed, causing a change of the current value and version. Therefore, any transaction based on an obsolete version is to be invalidated regardless of how correct they are \cite{DBLP:conf/blockchain2/MeirBMT19}.

There have been several workaround solutions to address this matter. For example, Composition-key can bypass MVCC validation because there is a new key generated for every transaction. However, this contradicts with the purpose of MVCC and introduces cumbersome key management. A retry mechanism is another workaround, such that client applications are notified of such conflicts and then retry submitting again. While applicable for some scenarios, this may not be suitable for others. There are also other techniques such as queuing transactions before submitting them to the blockchain. However, this does not benefit from the high throughput promised by HLF.
 
To sum up, in all test cases presented in table~\ref{table:caliperSet}, the MVCC conflict was present, which is an unpleasant issue that has to be addressed. We do not expect such an issue to appear when a few incidents are occurring less frequently. However, in certain extreme scenarios, we expect HLF to exhibit malfunctions due to MVCC conflicts. For example, when a very poorly performing cloud provider causes monitoring agents to report incidents every 1 second. Even when we attempt to increase the frequency of block batching, there is value in addressing and mitigating this malfunction ~\cite{Sukhwani2018}\cite{Yuan2020}.

\section{Conclusion and Future work}
This paper presents a blockchain-based SLA enforcement approach that aims to automate SLA incident management. It focuses on processing SLA violations and decision-making on the compliance status of cloud providers obligated by an SLA. It validates the dependability of the proposed enforcement logic based on accuracy diagnostic method. This paper also investigates HLF as underlying technology using Hyperledger Caliper. The evaluation showed that HLF performs well in terms of latency and possibly throughput. HLF can potentially serve the purpose of our proposed enforcement logic. However, since our approach expects high throughput from monitoring agents, the issue of HLF MVCC conflict has to be addressed. Otherwise, we cannot safely assume dependability. In future work, we will investigate a viable enhancement on HLF, or at least adapting our proposed approach in a way that does not compromise the performance of HLF.

\bibliographystyle{IEEEtran}
\bibliography{library}

\begin{thebibliography}{10}
\providecommand{\url}[1]{#1}
\csname url@samestyle\endcsname
\providecommand{\newblock}{\relax}
\providecommand{\bibinfo}[2]{#2}
\providecommand{\BIBentrySTDinterwordspacing}{\spaceskip=0pt\relax}
\providecommand{\BIBentryALTinterwordstretchfactor}{4}
\providecommand{\BIBentryALTinterwordspacing}{\spaceskip=\fontdimen2\font plus
\BIBentryALTinterwordstretchfactor\fontdimen3\font minus
  \fontdimen4\font\relax}
\providecommand{\BIBforeignlanguage}[2]{{%
\expandafter\ifx\csname l@#1\endcsname\relax
\typeout{** WARNING: IEEEtran.bst: No hyphenation pattern has been}%
\typeout{** loaded for the language `#1'. Using the pattern for}%
\typeout{** the default language instead.}%
\else
\language=\csname l@#1\endcsname
\fi
#2}}
\providecommand{\BIBdecl}{\relax}
\BIBdecl

\bibitem{Faniyi2015}
\BIBentryALTinterwordspacing
F.~Faniyi and R.~Bahsoon, ``{A Systematic Review of Service Level Management in
  the Cloud},'' \emph{ACM Comput. Surv.}, vol.~48, no.~3, pp. 1--27, dec 2015.
  [Online]. Available:
  \url{http://dl.acm.org/citation.cfm?doid=2856149.2843890}
\BIBentrySTDinterwordspacing

\bibitem{Alqahtani2019}
\BIBentryALTinterwordspacing
A.~Alqahtani, E.~Solaiman, P.~Patel, S.~Dustdar, and R.~Ranjan, ``{Service
  level agreement specification for end‐to‐end IoT application
  ecosystems},'' \emph{Softw. Pract. Exp.}, vol.~49, no.~12, pp. 1689--1711,
  dec 2019. [Online]. Available:
  \url{https://onlinelibrary.wiley.com/doi/abs/10.1002/spe.2747}
\BIBentrySTDinterwordspacing

\bibitem{Alqahtani2018a}
A.~Alqahtani, Y.~Li, P.~Patel, E.~Solaiman, and R.~Ranjan, ``{End-To-End
  Service Level Agreement Specification for IoT Applications},'' in \emph{Proc.
  - 2018 Int. Conf. High Perform. Comput. Simulation, HPCS 2018}.\hskip 1em
  plus 0.5em minus 0.4em\relax IEEE, oct 2018, pp. 926--935.

\bibitem{Alzubaidi2019}
A.~Alzubaidi, E.~Solaiman, P.~Patel, and K.~Mitra, ``{Blockchain-based SLA
  Management in the Context of IoT},'' \emph{IT Prof.}, 2019.

\bibitem{Mubeen2018}
\BIBentryALTinterwordspacing
S.~Mubeen, S.~A. Asadollah, A.~V. Papadopoulos, M.~Ashjaei, H.~Pei-Breivold,
  and M.~Behnam, ``{Management of Service Level Agreements for Cloud Services
  in IoT: A Systematic Mapping Study},'' \emph{IEEE Access}, vol.~6, pp.
  30\,184--30\,207, 2018. [Online]. Available:
  \url{https://ieeexplore.ieee.org/document/8016558/}
\BIBentrySTDinterwordspacing

\bibitem{Scheid2019}
E.~J. Scheid, B.~B. Rodrigues, L.~Z. Granville, and B.~Stiller, ``{Enabling
  Dynamic SLA Compensation Using Blockchain-based Smart Contracts},'' in
  \emph{2019 IFIP/IEEE Symp. Integr. Netw. Serv. Manag.}, 2019, pp. 53--61.

\bibitem{Rana2008}
\BIBentryALTinterwordspacing
O.~F. Rana, M.~Warnier, T.~B. Quillinan, F.~Brazier, and D.~Cojocarasu,
  ``{Managing Violations in Service Level Agreements},'' in \emph{Grid Middlew.
  Serv.}\hskip 1em plus 0.5em minus 0.4em\relax Boston, MA: Springer US, 2008,
  pp. 349--358. [Online]. Available:
  \url{http://link.springer.com/10.1007/978-0-387-78446-5{\_}23}
\BIBentrySTDinterwordspacing

\bibitem{Simundic2009}
\BIBentryALTinterwordspacing
A.-M. {\v{S}}imundi{\'{c}}, ``{Measures of Diagnostic Accuracy: Basic
  Definitions.}'' \emph{EJIFCC}, vol.~19, no.~4, pp. 203--11, jan 2009.
  [Online]. Available: \url{http://www.ncbi.nlm.nih.gov/pubmed/27683318
  http://www.pubmedcentral.nih.gov/articlerender.fcgi?artid=PMC4975285}
\BIBentrySTDinterwordspacing

\bibitem{Stallings2015}
W.~Stallings, \emph{{Foundations of modern networking: SDN, NFV, QoE, IoT, and
  Cloud}}.\hskip 1em plus 0.5em minus 0.4em\relax Addison-Wesley Professional,
  2016.

\bibitem{White2017}
\BIBentryALTinterwordspacing
G.~White, V.~Nallur, and S.~Clarke, ``{Quality of service approaches in IoT: A
  systematic mapping},'' \emph{J. Syst. Softw.}, vol. 132, pp. 186--203, oct
  2017. [Online]. Available:
  \url{https://www.sciencedirect.com/science/article/pii/S016412121730105X}
\BIBentrySTDinterwordspacing

\bibitem{Hussain2014}
\BIBentryALTinterwordspacing
W.~Hussain, F.~K. Hussain, and O.~K. Hussain, ``{Maintaining Trust in Cloud
  Computing through SLA Monitoring}.''\hskip 1em plus 0.5em minus 0.4em\relax
  Springer, Cham, 2014, pp. 690--697. [Online]. Available:
  \url{http://link.springer.com/10.1007/978-3-319-12643-2{\_}83}
\BIBentrySTDinterwordspacing

\bibitem{Cai2018}
W.~Cai, Z.~Wang, J.~B. Ernst, Z.~Hong, C.~Feng, and V.~C. Leung,
  ``{Decentralized Applications: The Blockchain-Empowered Software System},''
  \emph{IEEE Access}, vol.~6, pp. 53\,019--53\,033, sep 2018.

\bibitem{Androulaki2018}
\BIBentryALTinterwordspacing
E.~Androulaki, A.~Barger, V.~Bortnikov, C.~Cachin, K.~Christidis, A.~{De Caro},
  D.~Enyeart, C.~Ferris, G.~Laventman, Y.~Manevich, S.~Muralidharan, C.~Murthy,
  B.~Nguyen, M.~Sethi, G.~Singh, K.~Smith, A.~Sorniotti, C.~Stathakopoulou,
  M.~Vukoli{\'{c}}, S.~W. Cocco, and J.~Yellick, ``{Hyperledger Fabric: A
  Distributed Operating System for Permissioned Blockchains},'' jan 2018.
  [Online]. Available: \url{http://arxiv.org/abs/1801.10228
  http://dx.doi.org/10.1145/3190508.3190538}
\BIBentrySTDinterwordspacing

\bibitem{Ongaro2014}
D.~Ongaro and J.~Ousterhout, ``{In search of an understandable consensus
  algorithm},'' in \emph{2014 {\{}USENIX{\}} Annu. Tech. Conf.
  ({\{}USENIX{\}}{\{}ATC{\}} 14)}, 2014, pp. 305--319.

\bibitem{Wang2019}
\BIBentryALTinterwordspacing
S.~Wang, L.~Ouyang, Y.~Yuan, X.~Ni, X.~Han, and F.-Y. Wang,
  ``{Blockchain-Enabled Smart Contracts: Architecture, Applications, and Future
  Trends},'' \emph{IEEE Trans. Syst. Man, Cybern. Syst.}, pp. 1--12, 2019.
  [Online]. Available: \url{https://ieeexplore.ieee.org/document/8643084/}
\BIBentrySTDinterwordspacing

\bibitem{Magazzeni2017}
D.~Magazzeni, P.~Mcburney, and W.~Nash, ``{Validation and verification of smart
  contracts: A research agenda},'' \emph{Computer (Long. Beach. Calif).},
  vol.~50, no.~9, pp. 50--57, 2017.

\bibitem{Christidis2016}
\BIBentryALTinterwordspacing
K.~Christidis and M.~Devetsikiotis, ``{Blockchains and Smart Contracts for the
  Internet of Things},'' pp. 2292--2303, 2016. [Online]. Available:
  \url{http://ieeexplore.ieee.org/document/7467408/}
\BIBentrySTDinterwordspacing

\bibitem{Thakkar2018}
P.~Thakkar, S.~Nathan, and B.~Vishwanathan, ``{Performance Benchmarking and
  Optimizing Hyperledger Fabric Blockchain Platform},'' \emph{arXiv Prepr.
  arXiv1805.11390}, 2018.

\bibitem{Kuzlu2019}
M.~Kuzlu, M.~Pipattanasomporn, L.~Gurses, and S.~Rahman, ``{Performance
  analysis of a hyperledger fabric blockchain framework: Throughput, latency
  and scalability},'' in \emph{Proc. - 2019 2nd IEEE Int. Conf. Blockchain,
  Blockchain 2019}.\hskip 1em plus 0.5em minus 0.4em\relax Institute of
  Electrical and Electronics Engineers Inc., jul 2019, pp. 536--540.

\bibitem{Gorenflo2019}
C.~Gorenflo, S.~Lee, L.~Golab, and S.~Keshav, ``{FastFabric: Scaling
  Hyperledger Fabric to 20,000 Transactions per Second},'' in \emph{ICBC 2019 -
  IEEE Int. Conf. Blockchain Cryptocurrency}.\hskip 1em plus 0.5em minus
  0.4em\relax Institute of Electrical and Electronics Engineers Inc., may 2019,
  pp. 455--463.

\bibitem{Dinh2017a}
\BIBentryALTinterwordspacing
T.~T.~A. Dinh, J.~Wang, G.~Chen, R.~Liu, B.~C. Ooi, and K.-L. Tan,
  ``{BLOCKBENCH: A Framework for Analyzing Private Blockchains},'' in
  \emph{Proc. 2017 ACM Int. Conf. Manag. Data - SIGMOD '17}.\hskip 1em plus
  0.5em minus 0.4em\relax New York, New York, USA: ACM Press, 2017, pp.
  1085--1100. [Online]. Available:
  \url{http://dl.acm.org/citation.cfm?doid=3035918.3064033}
\BIBentrySTDinterwordspacing

\bibitem{Sukhwani2018}
\BIBentryALTinterwordspacing
H.~Sukhwani, N.~Wang, K.~S. Trivedi, and A.~Rindos, ``{Performance Modeling of
  Hyperledger Fabric (Permissioned Blockchain Network)},'' in \emph{2018 IEEE
  17th Int. Symp. Netw. Comput. Appl.}\hskip 1em plus 0.5em minus 0.4em\relax
  IEEE, nov 2018, pp. 1--8. [Online]. Available:
  \url{https://ieeexplore.ieee.org/document/8548070/}
\BIBentrySTDinterwordspacing

\bibitem{Chung2019}
\BIBentryALTinterwordspacing
G.~Chung, L.~Desrosiers, M.~Gupta, A.~Sutton, K.~Venkatadri, O.~Wong, and
  G.~Zugic, ``{Performance Tuning and Scaling Enterprise Blockchain
  Applications},'' dec 2019. [Online]. Available:
  \url{http://arxiv.org/abs/1912.11456}
\BIBentrySTDinterwordspacing

\bibitem{DBLP:conf/blockchain2/MeirBMT19}
\BIBentryALTinterwordspacing
H.~Meir, A.~Barger, Y.~Manevich, and Y.~Tock, ``{Lockless Transaction Isolation
  in Hyperledger Fabric},'' in \emph{{\{}IEEE{\}} Int. Conf. Blockchain,
  Blockchain 2019, Atlanta, GA, USA, July 14-17, 2019}.\hskip 1em plus 0.5em
  minus 0.4em\relax IEEE, 2019, pp. 59--66. [Online]. Available:
  \url{https://doi.org/10.1109/Blockchain.2019.00017}
\BIBentrySTDinterwordspacing

\bibitem{Yuan2020}
P.~Yuan, K.~Zheng, X.~Xiong, K.~Zhang, and L.~Lei, ``{Performance modeling and
  analysis of a Hyperledger-based system using GSPN},'' \emph{Comput. Commun.},
  vol. 153, pp. 117--124, mar 2020.

\end{thebibliography}

\end{document}